\documentclass[preprint,showpacs,preprintnumbers,amsmath,amssymb,nofootinbib]{revtex4}

\usepackage{graphicx}
\usepackage{dcolumn}
\usepackage{bm}
\usepackage{longtable}
\usepackage{latexsym}
\newcommand{\ket}[1]{|{#1}\rangle}
\newcommand{\bra}[1]{\langle {#1}|}

\newcommand{\axialjc}[2]{j\sp{#1}\sb{A{#2}}}

\newcommand{\vectorjc}[2]{j\sp{#1}\sb{V{#2}}}

\newcommand{\fpi}{f\sb{\pi}}
\newcommand{\mpi}{m\sb{\pi}}
\newcommand{\mnucl}{m\sb{N}}
\newcommand{\mrop}{m\sb{R}}
\newcommand{\mdel}{m\sb{\Delta}}

\begin{document}
\title{Chiral Symmetry and 
$N\sp{\ast}(1440)\rightarrow N\pi\pi$ Decay}

\author{H. Kamano}
 \email{kamano@ocunp.hep.osaka-cu.ac.jp}
\author{M. Morishita}
 \email{morisita@ocunp.hep.osaka-cu.ac.jp}
\author{M. Arima}
 \email{arima@ocunp.hep.osaka-cu.ac.jp}
\affiliation{
 Department of Physics, Osaka City University, Osaka 558-8585, Japan }

\date{\today}

\begin{abstract}
The $N\sp{\ast}(1440)\rightarrow N\pi\pi$ decay is studied by
making use of the chiral reduction formula. 
This formula suggests a scalar-isoscalar pion-baryon contact interaction
which is absent in the recent study of Hern{\'a}ndez et al.
The contact interaction is introduced into their model,
and is found to be necessary for the simultaneous 
description of $g\sb{RN\pi\pi}$ and the $\pi\pi$ and $\pi N$ invariant mass
distributions.
\end{abstract}


\maketitle
\section{\label{sec1}Introduction}
In the theoretical understanding of baryon spectra,
the Roper resonance $N\sp{\ast}(1440)$ is well known as
the controversial object.
Although the naive quark model assigns a radial excitation of 
constituent quarks to this resonance, its mass becomes so heavy that 
this interpretation does not work~\cite{Isg}.
In connection with this problem
there are some recent studies of baryon spectra
based on, for example, the chiral algebra~\cite{Bea02},
the pentaquark (two-diquark and antiquark) model~\cite{Jaf03}
and the chiral soliton model~\cite{Wei04}.
This is, however, still an open problem
about the structure of $N\sp{\ast}(1440)$.

Besides being characteristic of the baryon spectra,
the Roper resonance is important for the low energy hadron reactions.
In particular, two pion production 
reactions~\cite{Ose85,Ber95,Jen97,Gom96,Alv98}
and some nuclear reaction~\cite{Hir96} near threshold are sensitive to 
the $N\sp{\ast}(1440)\rightarrow N(\pi\pi)\sp{I=0}\sb{S\text{-wave}}$ decay,
in which the Roper resonance goes into a nucleon accompanied by two pions 
with \textit{S}-wave and isospin zero.
The importance of this decay is emphasized  
by Manley et~al.~\cite{Man92},
which Particle Data Table~\cite{PDG04} refers to. 
In their analysis, the two decay channels 
$N\sp{\ast}(1440)\rightarrow\Delta\pi$ and $N\epsilon$
are taken as the intermediate processes
of the two-pion decay of $N\sp{\ast}(1440)$.
Here the `$\epsilon$ meson' describes
the scalar-isoscalar $\pi\pi$ correlation 
which may correspond to the $\sigma$ meson.

Recently the $\pi\pi$ and $\pi N$ invariant mass 
distributions of $N\sp{\ast}(1440)\rightarrow N\pi\pi$ decay
are examined by Hern{\'a}ndez et~al~\cite{Her02}.
In their work the $\pi\pi$ re-scattering is explicitly considered 
to describe the scalar-isoscalar $\pi\pi$ correlation in
$N\sp{\ast}(1440)\rightarrow N(\pi\pi)\sp{I=0}\sb{S\text{-wave}}$ decay
instead of the $\epsilon$ propagation used in Manley's analysis.
Since there is no available data for the mass distributions,
Hern{\'a}ndez~et~al. considered Manley's analysis as ``experiment'',
and employed its result for fixing their model parameters.
Hernandez et al. well reproduced the ``experimental results''
of the invariant mass distributions.

Although their model is effective for the 
decay processes at the mass shell energy of $N\sp{\ast}(1440)$,
it does not work in the $\pi\pi N$ threshold region.
The coupling constant of the phenomenological Lagrangian
${\cal L}\sb{RN\pi\pi}=
g\sb{RN\pi\pi}\Psi\sp{\dag}\sb{N}\Psi\sb{R}\pi\sp{a}\pi\sp{a}$,
extracted from the $\pi\pi N$ threshold amplitude,
is much smaller than the one 
needed to explain the experimental data;
$|g\sb{RN\pi\pi}\sp{model}/g\sb{RN\pi\pi}\sp{expt}|=0.42$.
They concluded that some extra contributions are necessary 
in order to solve this problem.

Chiral symmetry is 
one of the guiding principles for studying the low energy hadron processes
such as $N\sp{\ast}(1440)\rightarrow N\pi\pi$ decay.
The systematic chiral expansion scheme is, however, difficult 
in the resonance region. 
As a result, phenomenological treatments
have often been applied to the processes including baryon excitations,
in which it is not clear whether chiral symmetry is appropriately reflected.

For the purpose of studying hadron reactions without missing chiral symmetry,
the chiral reduction formula developed by 
Yamagishi and Zahed~\cite{Yam96} is a powerful method.
This formula offers the Ward identity required by chiral symmetry
for the scattering amplitudes of any hadron processes involving the pion. 
We can discuss the general framework of pion induced reactions 
separately from the detail of specific models.

In this paper, we discuss the general structure of
$N\sp{\ast}(1440)\rightarrow N\pi\pi$ decay amplitude
by making use of the Ward identity derived from the chiral reduction formula.
Considering this general structure,
we find that a scalar-isoscalar contact interaction for
the pion-baryon vertex in
the $N\sp{\ast}(1440)\rightarrow N(\pi\pi)\sp{I=0}\sb{S\text{-wave}}$ 
amplitude is absent in Ref.~\cite{Her02}.
We bring this contact interaction 
in the model of Ref.~\cite{Her02},
and discuss its influence on 
the $N\sp{\ast}(1440)\rightarrow N\pi\pi$ decay.
The scalar-isoscalar contact interaction has two aspects
in its origin: one is due to the explicit chiral symmetry breaking 
and the other survives in the chiral limit.
We will see that the former part plays an important role to solve 
the problem in Ref~\cite{Her02}. 

We stress here that the scalar-isoscalar contact interaction 
is not intuitively introduced in our discussion, 
but naturally appears in the general structure of 
$N\sp{\ast}(1440)\rightarrow N\pi\pi$ amplitude
given by the Ward identity.
%
Therefore we should include this interaction
as long as there is no ad hoc reason to neglect it. 

This paper is organized as follows.
In Sec.~\ref{sec2}, we explain the general structure of the 
$N\sp{\ast}(1440)\rightarrow N\pi\pi$ amplitude
derived from the chiral reduction formula.
After a brief review of Ref.~\cite{Her02},
we introduce the scalar-isoscalar contact interaction
in the decay amplitude.
Taking account of this interaction,
we calculate the $\pi\pi$ and $\pi N$ invariant mass distributions
for the $N\sp{\ast}(1440)\rightarrow N\pi\pi$ decay.
We show our results and 
discuss the important role of the contact interaction in Sec.~\ref{sec3}.
Summary and conclusion are given in Sec.~\ref{sec4}.
\section{
\label{sec2}
Theoretical treatment of $N\sp{\ast}(1440)\rightarrow N\pi\pi$ decay
}
\begin{center}
\textbf{A. General structure of amplitude}
\end{center}
\begin{figure}
\includegraphics[width=3cm]{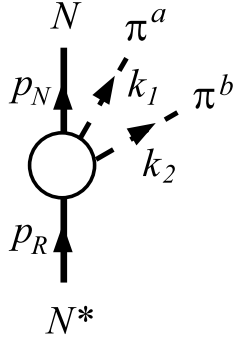}
\caption{
Kinematics of $N\sp{\ast}(1440)\rightarrow N\pi\pi$ decay.
Here $p\sb{R}$ ($p\sb{N}$) is the four momentum of the Roper (nucleon),
and $k\sb{1},k\sb{2}$ and $a,b$ are the four momenta and isospin indices
of pions. 
}
\label{fig1}
\end{figure}

Using the chiral reduction formula~\cite{Yam96}, we decompose the 
$N\sp{\ast}(1440)\rightarrow N\pi\pi$ amplitude $i{\cal T}$ as
(see Fig.~\ref{fig1}  for the kinematics of
the $N\sp{\ast}(1440)\rightarrow N\pi\pi$ decay)
\begin{equation}
i{\cal T}
=i{\cal T}\sb{S}+i{\cal T}\sb{V}+i{\cal T}\sb{AA}.
\label{eq1}
\end{equation}
Each term on the right hand side is given by
\begin{equation}
i{\cal T}\sb{S}=
-i\frac{\mpi\sp{2}}{\fpi}
\delta\sp{ab}
\bra{N(p\sb{N})}\hat{\sigma}(0)\ket{N\sp{\ast}(p\sb{R})},
\label{eq2}
\end{equation}
\begin{equation}
i{\cal T}\sb{V}=
-\frac{1}{2\fpi\sp{2}}(k\sb{1}-k\sb{2})\sp{\mu}\varepsilon\sp{abc}
\bra{N(p\sb{N})}\vectorjc{c}{\mu}(0)\ket{N\sp{\ast}(p\sb{R})},
\label{eq3}
\end{equation}
\begin{equation}
i{\cal T}\sb{AA}=
+\frac{1}{\fpi\sp{2}}k\sb{1}\sp{\mu}k\sb{2}\sp{\nu}
\int d\sp{4}xe\sp{ik\sb{1}x}
\bra{N(p\sb{N})}
T\sp{\ast}(\axialjc{a}{\mu}(x)\axialjc{b}{\nu}(0))
\ket{N\sp{\ast}(p\sb{R})},
\label{eq4}
\end{equation}
where $\hat{\sigma}$ and $\vectorjc{a}{\mu}$ are
the scalar density and the vector current, respectively, and
$\axialjc{a}{\mu}$ is the one-pion reduced axial current
which is obtained by extracting the one-pion component 
from the ordinary axial current.
The delta function 
$(2\pi)\sp{4}\delta\sp{(4)}(p\sb{R}-p\sb{N}-k\sb{1}-k\sb{2})$
for the momentum conservation is suppressed in these expressions.

${\cal T}\sb{S}$ appears owing to the explicit chiral symmetry breaking,
and vanishes in the chiral limit $\mpi\rightarrow 0$.
This term contributes only to
the $N\sp{\ast}(1440)\rightarrow N(\pi\pi)\sp{I=0}\sb{S\text{-wave}}$ 
amplitude because of the scalar-isoscalar nature of $\hat{\sigma}$.
${\cal T}\sb{AA}$ also contributes to
the $N\sp{\ast}(1440)\rightarrow N(\pi\pi)\sp{I=0}\sb{S\text{-wave}}$ 
amplitude. 
In contrast to ${\cal T}\sb{S}$,
this term does not vanish in the chiral limit.
The decay process in which baryon resonances appear
in the intermediate steps, such as 
$N\sp{\ast}(1440)\rightarrow \Delta\pi$, 
is included in ${\cal T}\sb{AA}$.
In the low energy region, ${\cal T}\sb{V}$ expresses
the amplitude of $N\sp{\ast}(1440)\rightarrow N\rho$ decay process
in which two pions are correlated in $P$-wave. 

Here we write explicit expressions of the  
$N\sp{\ast}(1440)\rightarrow N(\pi\pi)\sp{I=0}\sb{S\text{-wave}}$ amplitude.
By introducing the form factors, we have
\begin{equation}
i{\cal T}\left[
N\sp{\ast}(1440)\rightarrow N(\pi\pi)\sp{I=0}\sb{S\text{-wave}}
\right]=i{\cal T}\sb{S}+i{\cal T}\sb{AA}\sp{C},
\label{eq5}
\end{equation}
\begin{equation}
i{\cal T}\sb{S}=
i\delta\sp{ab}\frac{\sigma\sb{RN}(s\sb{\pi\pi})}{\fpi\sp{2}}
\Phi\sb{N}\sp{\dag}\Phi\sb{R}
\chi\sb{N}\sp{\dag}\chi\sb{R},
\label{eq6}
\end{equation}
\begin{equation}
i{\cal T}\sb{AA}\sp{C}=
-i\delta\sp{ab}(k\sb{1}\cdot k\sb{2})
\frac{F\sb{AA}(s\sb{\pi\pi})}{\fpi\sp{2}}
\Phi\sb{N}\sp{\dag}\Phi\sb{R}
\chi\sb{N}\sp{\dag}\chi\sb{R},
\label{eq7}
\end{equation}
where $s\sb{\pi\pi}=(k\sb{1}+k\sb{2})\sp{2}$ is the 
invariant mass square of the emitted two pions and 
$\Phi\sb{N}$ [$\Phi\sb{R}$] and $\chi\sb{N}$ [$\chi\sb{R}$]
are the isospin and spin Pauli spinors of the nucleon [Roper],
respectively. 
${\cal T}\sb{AA}\sp{C}$ is a part of ${\cal T}\sb{AA}$ contributing to
the $N\sp{\ast}(1440)\rightarrow N(\pi\pi)\sp{I=0}\sb{S\text{-wave}}$ decay. 
The nonrelativistic approximation is taken for the baryon states
in order to compare our results with those of Ref.~\cite{Her02}.
The form factor $\sigma\sb{RN}(s\sb{\pi\pi})$ vanishes in $\mpi\rightarrow 0$.
$F\sb{AA}(s\sb{\pi\pi})$ does not include the single baryon poles
owing to the notation 
of the $N\sp{\ast}(1440)\rightarrow N(\pi\pi)\sp{I=0}\sb{S\text{-wave}}$ decay.
\begin{center}
\textbf{B. Model of Hern{\'a}ndez et al.} 
\end{center}
\begin{figure}
\includegraphics[width=8cm]{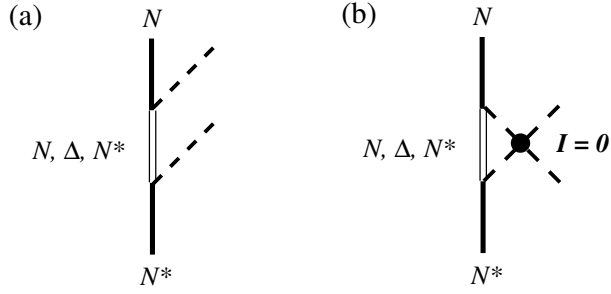}
\caption{
The `open' diagram (a), and the `closed' diagram (b)
for the $N\sp{\ast}(1440)\rightarrow N\pi\pi$ decay in Ref.~\cite{Her02}.
The pion-baryon vertices are given by the standard axial derivative coupling.
The blob in (b) expresses the $\pi\pi$ re-scattering.
}
\label{fig2}
\end{figure}
\begin{figure}
\includegraphics[width=10cm]{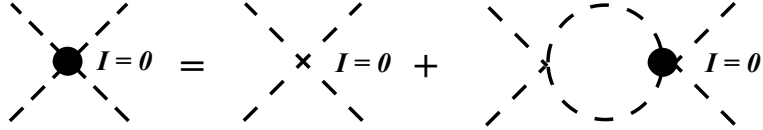}
\caption{
The diagrammatic expression of the $\pi\pi$ re-scattering mechanism 
in $I=0 $ channel.
}
\label{fig3}
\end{figure}

Now we proceed to a brief review of the model used in Ref.~\cite{Her02}.
This model consists of
the `open' and `closed' diagrams as shown in Fig.~\ref{fig2}.
The amplitudes are found in Ref.~\cite{Her02}
for the propagation of $\Delta$ as an intermediate baryon.
Similarly we can obtain the amplitudes also 
for the intermediate $N$ and $N\sp{\ast}(1440)$.
In this model,
the $N\sp{\ast}(1440)\rightarrow N(\pi\pi)\sp{I=0}\sb{S\text{-wave}}$ decay
takes place only through the closed diagrams.
The scalar-isoscalar correlation of the two pions is realized by 
the re-scattering mechanism based on the Lippmann-Schwinger equation
as shown in Fig.~\ref{fig3},
in which the tree level amplitude with $I=0$ is
provided by the lowest order chiral Lagrangian.
The $I=0$ $\pi\pi$ amplitude including the re-scattering effect
is\footnote{See Ref.~\cite{Oll97} 
for this solution on the basis of the on-shell tree amplitude.} 
\begin{equation}
t\sb{\pi\pi}\sp{I=0}(s\sb{\pi\pi})=
-\frac{6}{\fpi\sp{2}}\frac{s\sb{\pi\pi}-\mpi\sp{2}/2}
{1+(1/\fpi\sp{2})(s\sb{\pi\pi}-\mpi\sp{2}/2)G(s\sb{\pi\pi})}
\label{eq8}
\end{equation}
with the pion loop integral  
\begin{equation}
G(s\sb{\pi\pi})=
i\int \frac{d\sp{4}l}{(2\pi)\sp{4}}
\frac{1}{l\sp{2}-\mpi\sp{2}+i\varepsilon}
\frac{1}{(P-l)\sp{2}-\mpi\sp{2}+i\varepsilon},
\label{eq9}
\end{equation}
where $P\sp{2}=s\sb{\pi\pi}$.
Using the dimensional regularization scheme with a renormalization scale 
$\mu=1.2$~GeV, we obtain 
\begin{equation}
G(s\sb{\pi\pi})= 
\frac{1}{(4\pi)\sp{4}}
\left(-1+\ln\frac{\mpi\sp{2}}{\mu\sp{2}}
+\sigma\ln\frac{1+\sigma}{1-\sigma}-i\pi\sigma
\right),
\label{eq10}
\end{equation}
where $\sigma=\sqrt{1-(4\mpi\sp{2}/s\sb{\pi\pi})}$.

The amplitudes corresponding to the diagrams in Fig.~\ref{fig2}
can be classified as ${\cal T}\sb{S}$ and ${\cal T}\sb{AA}$.
The open diagram Fig.~\ref{fig2}~(a)
which expresses the decay processes through the baryon intermediate states
is included in ${\cal T}\sb{AA}$.
As for the closed diagram Fig.~\ref{fig2}~(b), we first note that
the decay amplitudes are proportional to $t\sb{\pi\pi}\sp{I=0}(s\sb{\pi\pi})$
(see Eq.~(8) in Ref.~\cite{Her02}).
Rewriting $t\sb{\pi\pi}\sp{I=0}(s\sb{\pi\pi})$ as 
\begin{equation}
t\sb{\pi\pi}\sp{I=0}(s\sb{\pi\pi})=
-\left(k\sb{1}\cdot k\sb{2}+\frac{3}{4}\mpi\sp{2}\right)
\frac{6}{\fpi\sp{2}}
\frac{2}{1+(1/\fpi\sp{2})(s\sb{\pi\pi}-\mpi\sp{2}/2)G(s\sb{\pi\pi})},
\label{eq11}
\end{equation}
we can see that each part of the decay amplitude proportional 
to $k\sb{1}\cdot k\sb{2}$ and $\mpi\sp{2}$ belongs to
${\cal T}\sp{C}\sb{AA}$ and ${\cal T}\sb{S}$,
respectively.
The phenomenological amplitudes of Ref.~\cite{Her02} are consistent
with the general structure based on the chiral reduction formula. 
\begin{center}
\textbf{C.~Scalar-isoscalar contact interaction}
\end{center}

In Ref.~\cite{Her02} it is only the $\pi\pi$ interaction that leads to
the $N\sp{\ast}(1440)\rightarrow N(\pi\pi)\sp{I=0}\sb{S\text{-wave}}$ decay.
The general expressions Eqs.~(\ref{eq5})-(\ref{eq7}) based on chiral symmetry,
however, also allow the scalar-isoscalar contact interaction for 
the pion-baryon vertex to be the source of this decay.

\begin{figure}
\includegraphics[width=8cm]{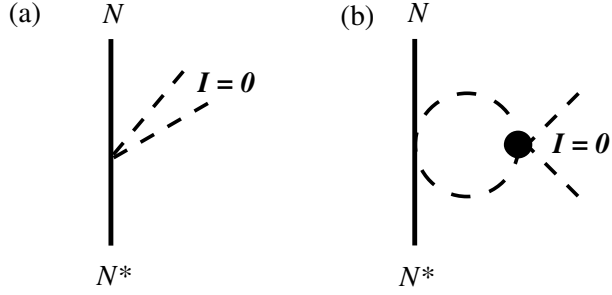}
\caption{
The scalar-isoscalar contact interaction suggested by the Ward identity;
(a) the tree diagram corresponding to the sum of 
Eqs.~(\ref{eq12})~and~(\ref{eq13}),
(b) the loop diagram including the $\pi\pi$ re-scattering vertex~(\ref{eq14}).
}
\label{fig4}
\end{figure}

The form factors in Eqs.~(\ref{eq6}) and (\ref{eq7})
are constants at tree level,
i.e. $\sigma\sb{RN}(s\sb{\pi\pi}), F\sb{AA}(s\sb{\pi\pi})\rightarrow 
\sigma\sb{RN}, F\sb{AA}$, and we obtain
\begin{equation}
i{\cal T}\sb{S}\rightarrow
i\delta\sp{ab}\frac{\sigma\sb{RN}}{\fpi\sp{2}}
\Phi\sb{N}\sp{\dag}\Phi\sb{R}
\chi\sb{N}\sp{\dag}\chi\sb{R},
\label{eq12}
\end{equation}
\begin{equation}
i{\cal T}\sb{AA}\sp{C}\rightarrow
-i\delta\sp{ab}(k\sb{1}\cdot k\sb{2})\frac{F\sb{AA}}{\fpi\sp{2}}
\Phi\sb{N}\sp{\dag}\Phi\sb{R}
\chi\sb{N}\sp{\dag}\chi\sb{R}.
\label{eq13}
\end{equation}
They are the scalar-isoscalar contact interactions for the pion-baryon vertex
in the $N\sp{\ast}(1440)\rightarrow N(\pi\pi)\sp{I=0}\sb{S\text{-wave}}$ 
amplitude.
It is natural to take account of these contact interactions
in phenomenological treatments
because there is no reason to discard these interactions.
There is a difference in momentum dependence
of Eqs.~(\ref{eq12})~and~(\ref{eq13}) 
arising from the origin of each amplitude in the chiral structure.
Combining the contact interactions with the $\pi\pi$ re-scattering mechanism
of Ref.~\cite{Her02},
we obtain a new amplitude depicted in Fig.~\ref{fig4},
\begin{equation}
i{\cal T}\sb{new}=
i\frac{\delta\sp{ab}}{\fpi\sp{2}}
{\bm (}\sigma\sb{RN}-(k\sb{1}\cdot k\sb{2})F\sb{AA}{\bm )}
\left(1+\frac{1}{6}G(s\sb{\pi\pi})
t\sb{\pi\pi}\sp{I=0}(s\sb{\pi\pi})\right)
\Phi\sb{N}\sp{\dag}\Phi\sb{R}
\chi\sb{N}\sp{\dag}\chi\sb{R}.
\label{eq14}
\end{equation}
We calculate the $\pi\pi$ and $\pi N$ invariant mass distributions
of $N\sp{\ast}(1440)\rightarrow N\pi\pi$ decay including 
Eq.~(\ref{eq14}) in the phenomenological amplitude of Ref.~\cite{Her02}. 

Before ending this section, we note that ${\cal T}\sb{V}$ is not considered 
in the following calculation because this term is irrelevant to 
the $N\sp{\ast}(1440)\rightarrow N(\pi\pi)\sp{I=0}\sb{S\text{-wave}}$ decay,
which is also absent in Refs.~\cite{Man92,Her02}.
\section{\label{sec3}Results and Discussions}
\begin{table}
\caption{
The value of constants in our calculation.
The mass and total decay width of the Roper resonance 
is taken from Ref.~\cite{Man92}.
As for the coupling constants,
we use the same values as Ref.~\cite{Her02}. 
See Ref.~\cite{Her02} also for the explicit form of the interaction Lagrangian.
}
\label{tab1}
\begin{ruledtabular}
\begin{tabular}{cccc}
Masses and Width&(MeV)	&Constants		&	\\ \hline
$\mpi$	&139	&$f\sb{\pi NN}$		&0.95	\\
$\mnucl$&939	&$f\sb{\pi N\Delta}$	&2.07	\\
$\mdel$	&1232	&$f\sb{\pi NR}$		&0.40	\\
$\mrop$	&1462   &$f\sb{\pi RR}$		&0.95	\\
$\Gamma\sb{R}$&391  	&$f\sb{\pi}$		&92.4 MeV
\end{tabular}
\end{ruledtabular}
\end{table}

In this section, we show our results for the $\pi\pi$ and $\pi N$
invariant mass distributions of $N\sp{\ast}(1440)\rightarrow N\pi\pi$ decay
calculated in the rest frame of the Roper resonance.
We take $\Lambda$, $\sigma\sb{RN}$, and $F\sb{AA}$
as free parameters, which are 
the cutoff to regularize the closed diagram,
and the form factors introduced in the last section, respectively.
The values of these parameters are chosen so that we reproduce 
not only the mass distributions and 
the $N\sp{\ast}(1440)\rightarrow N\pi\pi$ decay width 
$\Gamma=152$ MeV estimated by using Manley's approach,
but also $g\sp{expt}\sb{RN\pi\pi}= 1.6\times 10\sp{-2}~\text{MeV}\sp{-1}$.
As for the coupling constant of the $R\Delta\pi$ interaction 
$f\sb{R\Delta\pi}$, we use $f\sb{R\Delta\pi}=1.1$
so that we have the same contribution of the open diagrams
as Ref.~\cite{Her02}.
In Table~\ref{tab1} we summarize the values of masses and other constants 
which are fixed throughout this calculation.
\begin{figure}
\includegraphics[width=7cm]{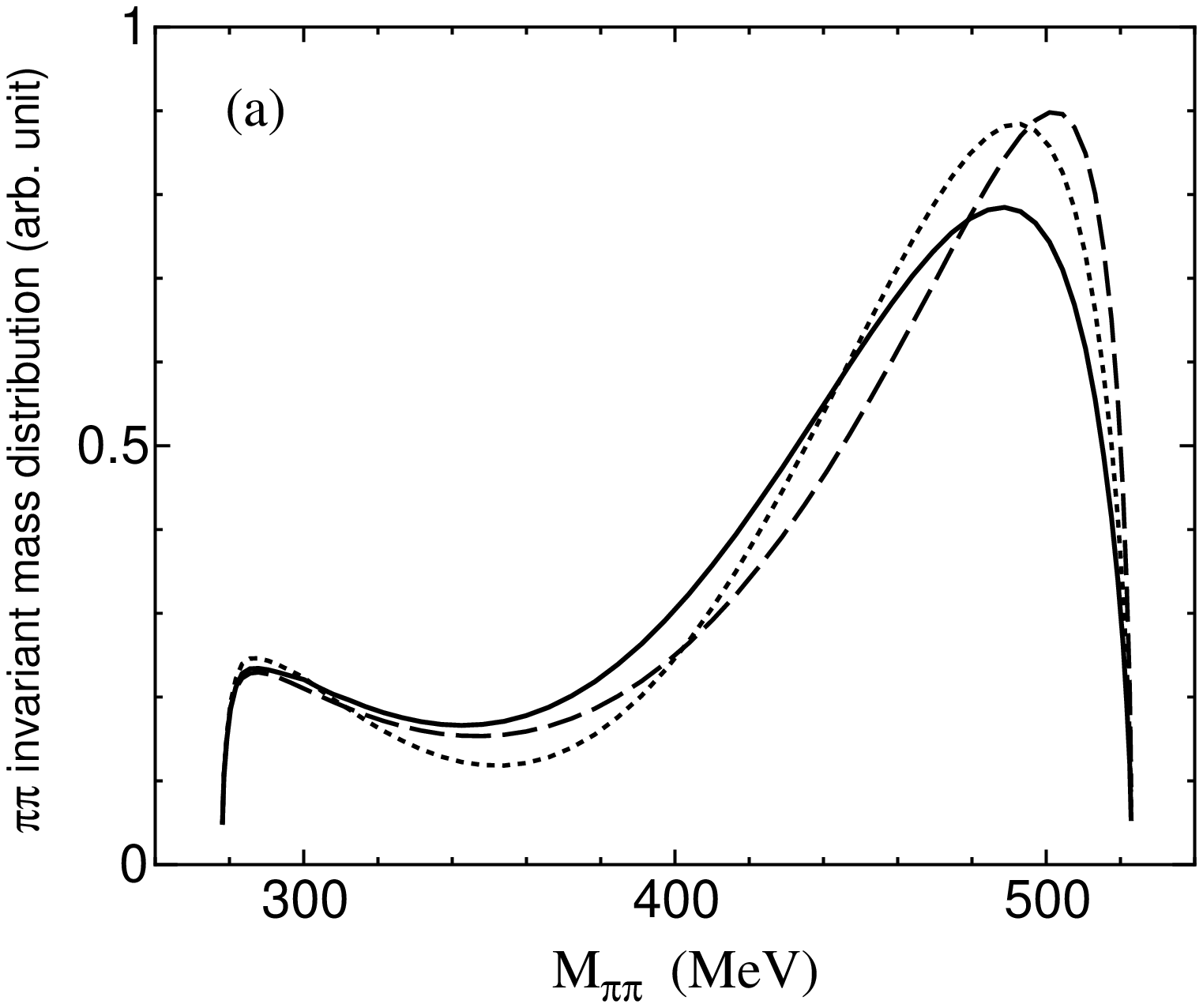}
\includegraphics[width=7cm]{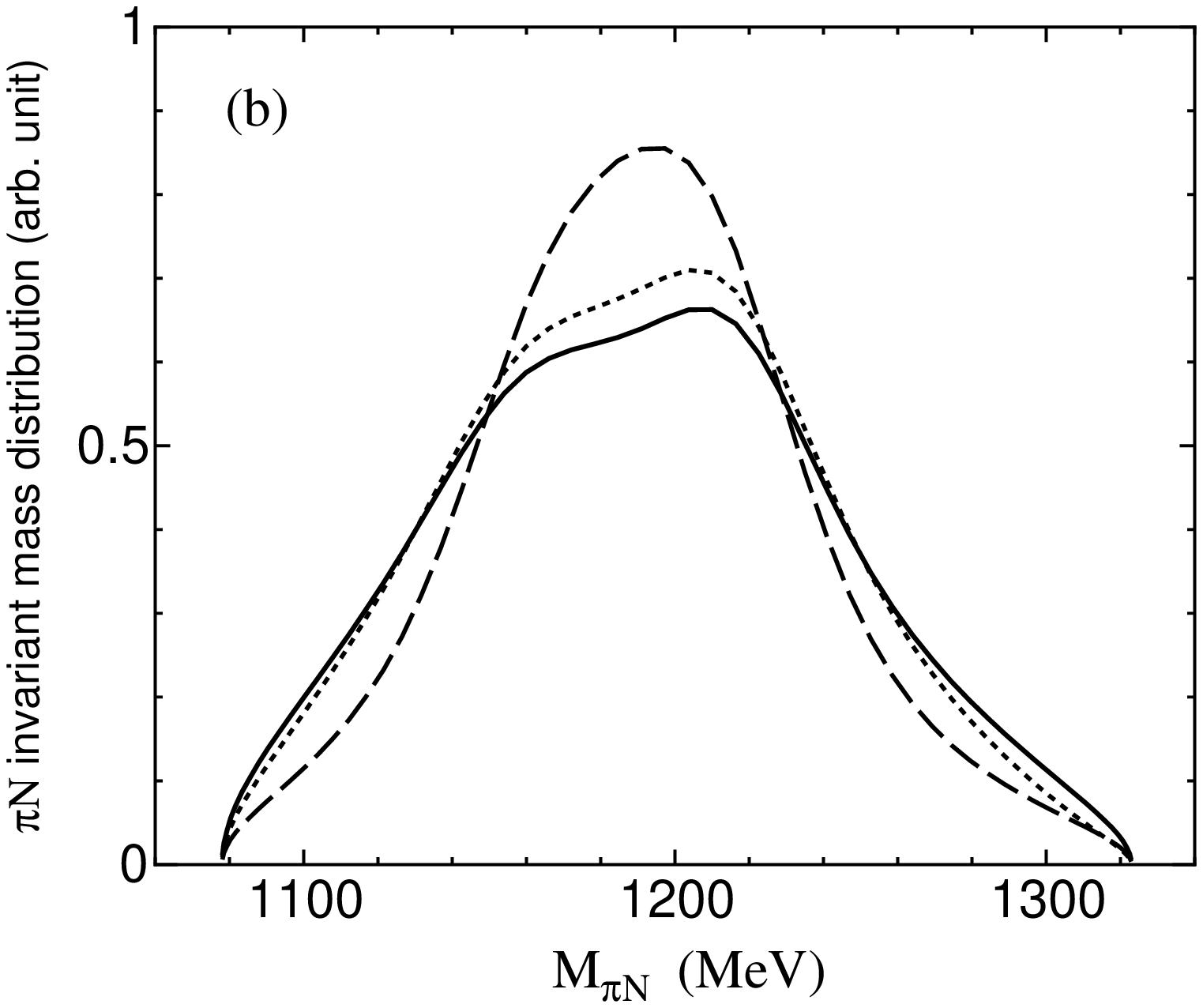}
\caption{
The (a) $\pi\pi$ and (b) $\pi N$
invariant mass distributions in our full calculation.
The $M\sb{\pi\pi}$ and $M\sb{\pi N}$ are the $\pi\pi$ and 
$\pi N$ invariant masses, respectively.
Our result is shown by the solid line.
The ``experimental result'' (Manley's approach)
and the result of Ref.~\cite{Her02} are shown by the
dashed and dotted line, respectively.
}
\label{fig5}
\end{figure}
\begin{center}
\textbf{A.~Full results}
\end{center}

Fig.~\ref{fig5} shows the mass distributions calculated 
by using ${\cal T}\sb{new}$ and the amplitudes corresponding to
the open and closed diagrams (the full calculation).
The parameters are $\sigma\sb{RN}=157~\text{MeV}$, 
$F\sb{AA}=1.97\times 10\sp{-3}~\text{MeV}\sp{-1}$,
and $\Lambda=700~\text{MeV}$.
We also display the results of Ref.~\cite{Her02} for comparison.
Both calculations are able to reproduce the ``experimental results''
of the $\pi\pi$ and $\pi N$ invariant mass distributions well.
Note that our model succeeded in reproducing $g\sp{expt}\sb{RN\pi\pi}$ 
together with the mass distributions and the decay width, 
while $|g\sp{model}\sb{RN\pi\pi}/g\sp{expt}\sb{RN\pi\pi}|=0.42$
in Ref.~\cite{Her02}.
Because our $\Lambda$ and $f\sb{R\Delta\pi}$ take the same values
as those of Ref.~\cite{Her02}, the difference between 
these two calculations is purely due to the amplitude Eq.~(\ref{eq14}).
Taking account of ${\cal T}\sb{new}$,
we obtain the consistent results in our calculation 
for the mass distributions and $g\sb{RN\pi\pi}$.
\begin{center}
\textbf{B.~$\sigma\sb{RN}$ and $F\sb{AA}$}
\end{center}
\begin{figure}
\includegraphics[width=7cm]{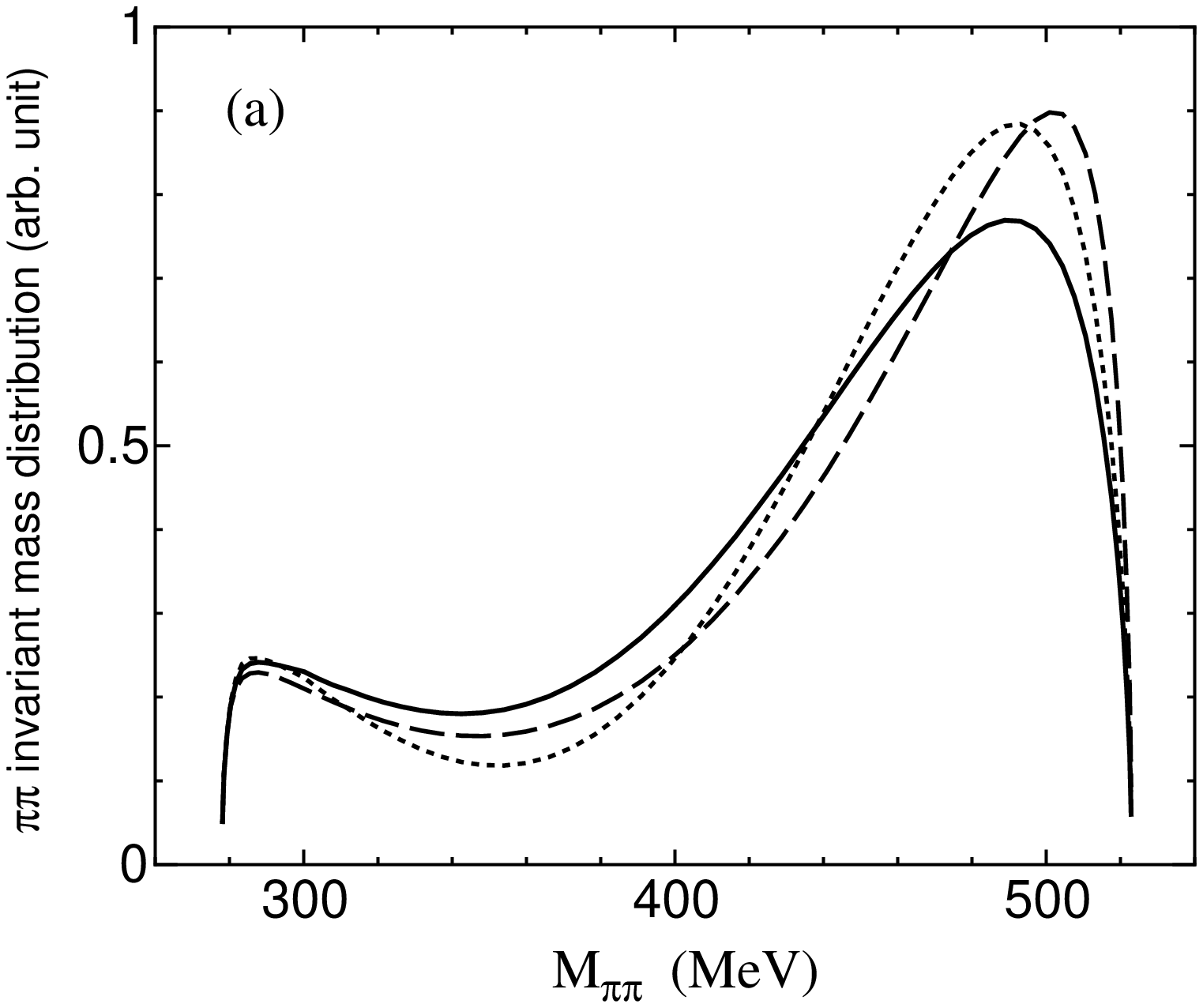}
\includegraphics[width=7cm]{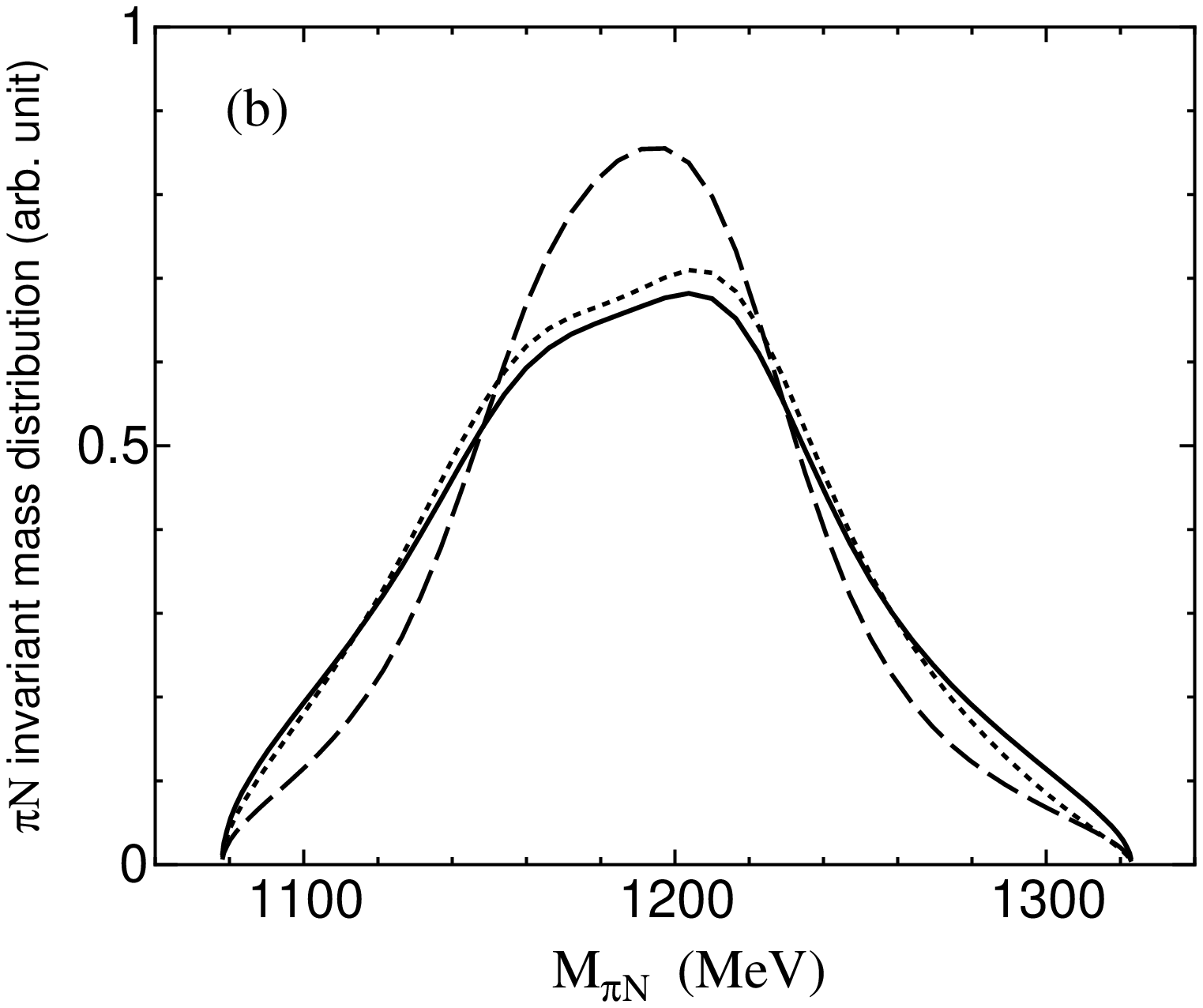}
\caption{
Same as Fig.~\ref{fig5}, but for the case of $F\sb{AA}=0$.
}
\label{fig6}
\end{figure}

We further examine the contribution of ${\cal T}\sb{new}$ on the
mass distributions.
This amplitude consists of two terms proportional to
$\sigma\sb{RN}$ and $F\sb{AA}$.
The former is connected with the explicit breaking of chiral symmetry
in the pion-baryon contact interaction [Eq.~(\ref{eq12})],
and the latter is derived from
the scalar-isoscalar combination of two axial vector currents 
[Eq.~(\ref{eq13})].

Now we estimate each contribution of these terms separately.
First, we set $F\sb{AA}$ 
to be zero and consider only the term proportional to $\sigma\sb{RN}$
with respect to ${\cal T}\sb{new}$. 
We obtain $\sigma\sb{RN}=160~\text{MeV}$ and $\Lambda=460~\text{MeV}$,
and also use $f\sb{R\Delta\pi}=1.1$.
We find that this calculation reproduces the mass distributions 
(Fig.~\ref{fig6}) and $g\sb{RN\pi\pi}$ as well as the full calculation. 
In this case $\Lambda$ becomes considerably small,
while $\sigma\sb{RN}$ takes almost the same value as before.

Next we set $\sigma\sb{RN}$ to be zero and consider the term
proportional to $F\sb{AA}$ in ${\cal T}\sb{new}$.
In this case we can not find acceptable values for the parameters
$F\sb{AA}$ and $\Lambda$ reproducing the mass distributions and 
$g\sb{RN\pi\pi}$ even though we treat $f\sb{R\Delta\pi}$ as a free parameter.
If we draw the mass distributions 
which is similar to Figs.~\ref{fig5}~and~\ref{fig6},
we obtain $g\sb{RN\pi\pi}$ far from
$g\sp{expt}\sb{RN\pi\pi}=1.6\times 10\sp{-2}~\text{MeV}\sp{-1}$.
This is also the case for Ref.~\cite{Her02}
as explained in the introduction.
These results imply that 
the term proportional to $\sigma\sb{RN}$ is necessary in reproducing
the mass distributions and $g\sb{RN\pi\pi}$ \textit{simultaneously}. 
\begin{center}
\textbf{C. Phenomenological meaning of closed diagrams}
\end{center}
\begin{figure}
\includegraphics[width=7cm]{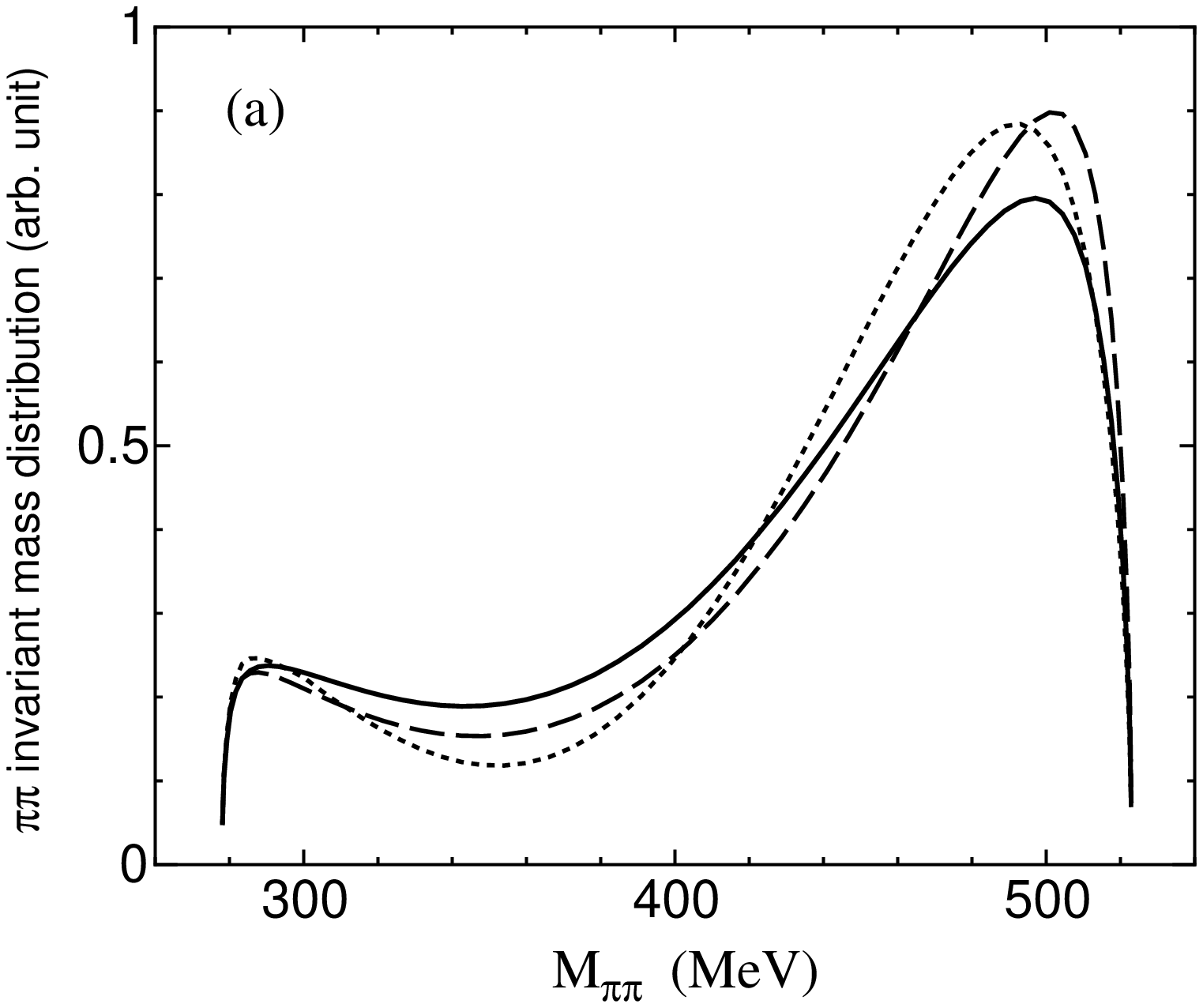}
\includegraphics[width=7cm]{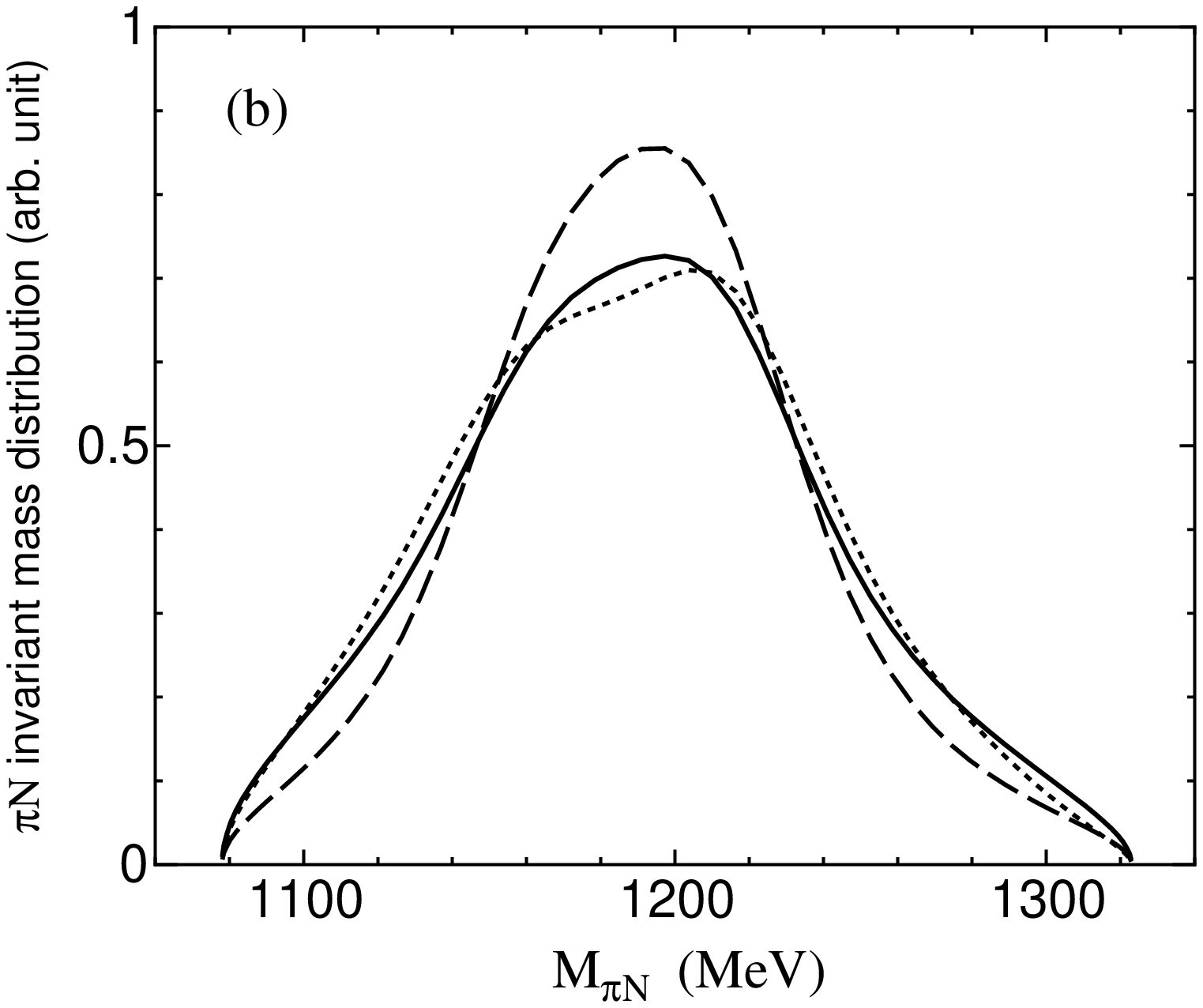}
\caption{
Same as Fig.~\ref{fig5}, but for the case without the closed diagrams.
}
\label{fig7}
\end{figure}

We consider the phenomenological meaning of the closed diagram.
There is no doubt that we should take the $\pi\pi$ correlation
into account when we study the two-pion decay of the Roper resonance.
We know that, however, there exist several ambiguities in 
the closed diagram; that is, this diagram depends highly on 
experimentally unknown cutoffs and coupling constants.
Is it possible to avoid using
the closed diagram for the phenomenological understanding 
of the $N\sp{\ast}(1440)\rightarrow N(\pi\pi)\sp{I=0}\sb{S\text{-wave}}$ decay?

We try to discuss the case that 
includes all diagrams explained above but the closed diagrams. 
In this case we also obtain the 
parameter set so as to reproduce the mass distributions (Fig.~\ref{fig7})
and $g\sb{RN\pi\pi}$.
The parameters become $\sigma\sb{RN}=177~\text{MeV}$ and 
$F\sb{AA}=-1.22\times 10\sp{-3}~\text{MeV}\sp{-1}$.
In comparison with the parameters in our full calculation,
the change in the value of $\sigma\sb{RN}$ is small, 
while $F\sb{AA}$ changes its sign from positive to negative.

This result shows that ${\cal T}\sb{new}$ effectively represents
the contribution of the closed diagrams 
by tuning $\sigma\sb{RN}$ and, especially $F\sb{AA}$.
\section{\label{sec4}Conclusions}
We have shown that the amplitude proportional to $\sigma\sb{RN}$,
which is connected with the explicit chiral symmetry breaking
of pion-baryon interaction, is necessary for the simultaneous 
description of $g\sb{RN\pi\pi}$ and the $\pi\pi$ and $\pi N$ invariant 
mass distributions.
Furthermore we have also shown that our amplitude derived from the 
contact interactions with $\sigma\sb{RN}$ and $F\sb{AA}$ are effectively
substituted for the amplitude corresponding to the closed diagrams of 
Ref.~\cite{Her02}.

A naive application of our experience at the low energy,
in which the explicit breaking of chiral symmetry can be 
neglected for the phenomenological analysis,
does not work for the present case at resonance energy.
And, unless there is theoretical and/or experimental evidence, 
the closed diagram including baryon resonances is not prior to 
the contact interaction.
In fact, we see that the closed diagram does not play an essential
role for the phenomenological understanding of this decay.
It will be desirable to study the $N\sp{\ast}(1440)\rightarrow N\pi\pi$ decay
without using the closed diagram including some ambiguities for the
baryon dynamics.

Motivated by the chiral reduction formula, we have introduced
the contact interactions, and have obtained satisfactory results.
The general discussion based on the chiral reduction formula
tells us an effective way of the 
phenomenological treatment of hadron reactions.
\begin{acknowledgments}
The authors acknowledge the nuclear theory group of Osaka-City University
for useful discussions.
\end{acknowledgments}
%
%
%
%
%
%

%
%
%
%
\end{document}